\documentclass[draft]{agujournal2018}
\usepackage{apacite}
\usepackage[export]{adjustbox}
\usepackage{url}
\usepackage{amsmath}
\usepackage{amssymb}

\draftfalse

\journalname{Geophysical Research Letters}
\usepackage{gensymb}
\usepackage{amsmath}

\begin{document}

%
%


\title{Machine-Learning-Driven New Geologic Discoveries at Mars Rover Landing Sites: Jezero and NE Syrtis}

%
%




\authors{Murat Dundar \affil{1}, Bethany L. Ehlmann \affil{2,3}, Ellen K. Leask \affil{2}}

 \affiliation{1}{Computer \& Information Science Dept., Indiana University-Purdue University, Indianapolis, IN, USA,}
  \affiliation{2}{Div. of Geological \& Planetary Sciences, California Institute of Technology, Pasadena, CA, USA,}
   \affiliation{3}{Jet Propulsion Laboratory, California Institute of Technology, Pasadena, CA, USA}





\correspondingauthor{Murat Dundar}{mdundar@iupui.edu}




\begin{keypoints}
\item Machine learning can be highly effective in exposing tiny outcrops of rare phases in CRISM 
\item A new hydrated iron oxide phase, elsewhere on Mars attributed to akageneite, is detected in NE Syrtis and Jezero
\item Al clays, jarosite, chlorite-smectite, and hydrated silica are reported in Jezero
\end{keypoints}

%
%


\begin{abstract}

A hierarchical Bayesian classifier is trained at pixel scale with spectral data from the CRISM (Compact Reconnaissance Imaging Spectrometer for Mars) imagery. Its utility in detecting rare phases is demonstrated with new geologic discoveries near the Mars-2020 rover landing site. Akaganeite is found in sediments on the Jezero crater floor and in fluvial deposits at NE Syrtis. Jarosite and silica are found on the Jezero crater floor while chlorite-smectite and Al phyllosilicates are found in the Jezero crater walls. These detections point to a multi-stage, multi-chemistry history of water in Jezero crater and the surrounding region and provide new information for guiding the Mars-2020 rover's landed exploration. In particular, the akaganeite, silica, and jarosite in the floor deposits suggest either a later episode of salty, Fe-rich waters that post-date Jezero delta or groundwater alteration of portions of the Jezero sedimentary sequence.  

\end{abstract}

%
%

%


%
%
%
%

\section{Introduction}

Hyperspectral data collected by the Compact Reconnaissance Imaging Spectrometer for Mars (CRISM) aboard the Mars Reconnaissance Orbiter have proven instrumental in the discovery of a broad array of aqueous minerals on the surface of Mars since 2006 \citep{crism_sp,murchie2009synthesis,crism_sp2}. Although these data have revolutionized our understanding of the planet, existing geologic discoveries are mostly limited to common mineral phases that occur frequently and with relatively larger spatial extent. Secondary or accessory phases on Mars that occur infrequently or at low abundances in only a few locales are important for a more complete and accurate interpretation of the geologic processes that formed these phases, which in turn is critical for resolving questions of Mars's changing habitability. For example, specific minerals such as alunite and jarosite (acidic), serpentine (alkaline, reducing), analcime (alkaline, saline), prehnite (200 \degree C $<$ temperature $<$ 400 \degree C), and perhaps phases yet to be discovered, serve as direct environmental indicators of the geochemistry of waters on the Mars surface. Moreover, the identification of rare phases, even in just a few pixels, enables characterization of the mineral assemblages within a geologic unit, which are critical for identifying the thermodynamic conditions and fluid composition during interactions of rocks with liquid water. 

Isolation and discovery of accessory mineral phases is challenging due to the systematic artifacts, random noise, and other limitations of an aging instrument affecting more recently collected CRISM images. The most common spectral mineral-identification method involves finding the ratio of the average spectra from two regions along-track in the image, where the numerator is the spectrum from the area of interest and the denominator is the spectrum derived from a spectrally homogeneous bland region. Summary parameters derived from key absorption bands are used to identify candidate regions for the numerator and denominator. Although summary parameters have been quite  effective for detecting common phases with relatively larger spatial extent, distinctive absorption bands useful for detecting rare accessory phases cannot be reliably recovered by summary parameters due to two main reasons. First, rare phases span a limited number of nearby but not necessarily contiguous pixels in an image, which makes spectral averaging less useful compared to common phases in eliminating random noise. Second, key absorption bands of rare secondary minerals can occur at wavelengths close to the key absorption bands of common phases in the image. The 6.55 $nm$ increments between two channels in CRISM offer enough spectral resolution to differentiate between such primary and secondary phases in ideal conditions. However, considering the practical limitations of CRISM data and the occurrence of phases in mixtures, such a distinction may not be possible without exploiting the spectral data in its entirety and identifying less obvious spectral features characterizing these phases in a given locale. 

As part of our ongoing efforts to implement machine learning methods to fully automate mineral discovery in CRISM data, we have previously reported dozens of new jarosite and alunite detections across Mars \citep{dundar:2016a,ehlmann_2015a} and have identified a previously unknown CRISM artifact that mimics the characteristics of real mineral absorption at 2.1 $\mu m$ range that could have significant implications in the search for perchlorate \citep{leask2018new}. Here, we present technical details of our hierarchical Bayesian model and demonstrate its utility by reporting new rare discoveries from the NE Syrtis area and Jezero crater.  Jezero crater and the Syrtis are of high interest as regions where the Mars-2020 rover will conduct its in situ exploration and as some of the most dust-free and ancient areas where strata are well-exposed for study of Mars’ geologic history. Prior studies of Jezero crater and its watershed have focused primarily on the Fe/Mg smectite clays and carbonates that make up deltaic and crater floor deposits \citep{ehlmann2008orbital,ehlmann_2009a,goudge2015assessing}. Here, we focus on identification of small, rare phases to inform the geologic history of the crater in both the crater floor lake sediments, wallrock of Jezero, and surrounding region. The region is a well-suited proving ground for the proposed Bayesian model because of its mineral diversity, excellent image availability, and high relevance for Mars exploration.

\section{Methods}

\subsection{Image datasets}
I/F data are used as the primary source of data in this study. I/F data are derived by dividing surface radiance by solar irradiance. Radiance data are only used for ruling out certain artifacts during verification process. Simple atmospheric and photometric corrections are applied to all images using CRISM Analysis Toolkit \cite{cat,murchie2009compact}. Only spectral channels that cover the spectral region from 1.0 to 2.6$\mu$m (248 channels) are used in this study. 

Geographically projected CRISM data were co-registered with high resolution Context Imager (CTX) \citep{malin2007context} and HiRISE \citep{mcewen2007mars} image data. The CTX global mosaic was used as the basemap for examination of morphology \citep{dickson2018global}, and standard pipelines for producing local digital elevation models were produced using Caltech’s Murray Laboratory pipeline, which utilizes the Ames stereo pipeline \citep{beyer2018ames}. Our methods have been developed in multiple phases as described in the following sections. 

\subsection{Creating a training library of spectral patterns by unsupervised learning and visual classification}
\label{sec:lib}

Over fifty CRISM images from the Nili Fossae and Mawrth Vallis regions were processed by a nonparametric Bayesian clustering technique \cite{i2gmm}. This method generates a few hundred spectra per image processed,  which are visually inspected and classified to create a spectral training library. This unsupervised learning approach is not only very computational but also requires a tedious task of manually assigning extracted spectra to classes. Nonetheless, this step is needed toward fully automating mineral discovery. In the second phase, the training library collected in this phase is used to implement two models: a bland pixel scoring function for column-wise ratioing and a classifier model that operates on the ratioed data to render mineral classification. 
Both the scoring function and the classifier uses our two-layer Bayesian Gaussian mixture model. 

\subsection{Two-layer Bayesian Gaussian Mixture Model}
\label{sec:i2gmm}
\begin{figure}[p]
 \centering
 \includegraphics[width=\textwidth]{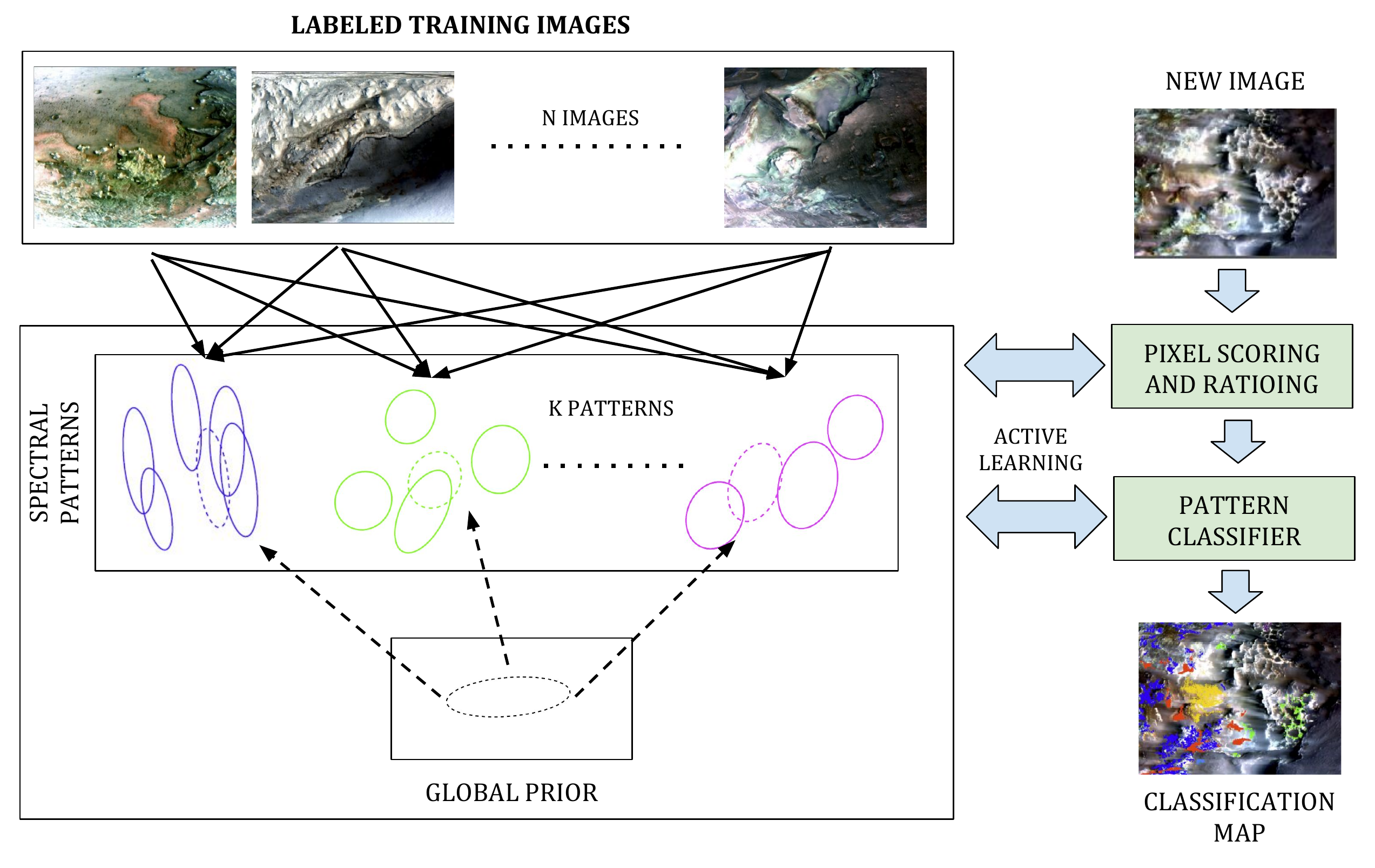}
 \caption{Two-layer Bayesian Gaussian Mixture Model Training and  Classification}
 \label{illustration}
  \end{figure}

Note that true distributions of spectral patterns in the training library are not known. Different instances of the same pattern detected across different images  exhibit varying spectral properties due to differences in atmospheric effects and viewing geometry as well as inherent differences in surface material spectral properties. Our two-layer Gaussian mixture model uses one mixture model for each spectral pattern in the lower layer. Herein, a spectral pattern might represent a  mineral phase, a known artifact, a bland pixel category, a common mixed phase, or an unidentified pattern.  The number of components in a mixture model for a given pattern is determined by the number of images in which that pattern occurs as the model introduces one Gaussian component for every image the pattern is detected. For example, out of 330 images available in our current training library jarosite exists in 44 of them, which implies that there are 44 observed instances of the jarosite pattern (``instance" refers to an occurrence in an image, which can be one or several pixels). The model introduces a Gaussian component for each instance to spectrally model the jarosite pixels corresponding to that instance. Gaussian components corresponding to the same spectral pattern are regulated by a shared local prior and local priors associated with each pattern are in turn modeled by a global prior. In this context the local prior can be thought of as the estimate for the true distribution of the corresponding pattern and the global prior can be interpreted as a template for all viable spectral patterns. This two-layer hierarchical model (illustrated in Figure \ref{illustration}) offers extreme flexibility and robustness for modeling pattern distributions. The lower layer models spectral variations of the same pattern across images whereas the upper layer models spectral variations across patterns. 
More specifically, we use the following generative model to fit spectral data available in our training library.
\begin{align}
\text{Data model: } \boldsymbol{x_{ijk}} \sim  N(\boldsymbol{\mu_{jk}}, \Sigma_{k})\\ \text{Local prior: } \boldsymbol{\mu}_{jk} \sim  N(\boldsymbol{\mu}_{k}, \Sigma_{k}\kappa_{1}^{-1}) \\ \text{Global prior: } \boldsymbol{\mu_{k}}  \sim  N(\boldsymbol{\mu_{0}}, \Sigma_{j}\kappa_{0}^{-1}), \Sigma_{k} \sim & W^{-1}(\Sigma_{0},m)
\label{eq:i2gmm}
\end{align}
where $k$, $j$, and $i$ are indices used to indicate true patterns, their observed instances, and individual pixels, respectively. $W^{-1}(\Sigma_{0},m)$ denotes the inverse Wishart distribution with scale matrix $\Sigma_{0}$ and degrees of freedom $m$. This model assumes that pixels $\boldsymbol{x_{ijk}}$ are distributed according to a Gaussian distribution with mean $\boldsymbol{\mu_{jk}}$ and covariance matrix $\Sigma_{k}$. Each true pattern is characterized by the parameters $\boldsymbol{\mu_{k}}$ and $\Sigma_{k}$. The parameter $\boldsymbol{\mu_{0}}$ is the mean of the Gaussian prior defined over the mean vectors of true patterns, $\kappa_{0}$ is a scaling constant that adjusts the dispersion of the centers of true patterns around $\boldsymbol{\mu_{0}}$. A smaller value for $\kappa_{0}$ suggests that pattern means are expected to be farther apart from each other whereas a larger value suggests they are expected to be closer. On the other hand, $\Sigma_{0}$ and $m$ dictate the expected shape of the pattern covariance matrices, as under the inverse Wishart distribution assumption the expected covariance is $E(\Sigma|\Sigma_{0},m)=\frac{\Sigma_{0}}{m-d-1}$, where $d$ denotes the number of channels used. The minimum feasible value of $m$ is equal to $d+2$, and the larger the $m$ is the less individual covariance matrices will deviate from the expected shape. The $\kappa_{1}$ is a scaling constant that adjusts the dispersion of the means of observed pattern instances around the centers of their corresponding true patterns. A larger $\kappa_{1}$ leads to smaller variations in instance means with respect to the means of their corresponding true pattern, suggesting small variations among observed instances of the pattern. On the other hand, a smaller $\kappa_{1}$ dictates larger variations among instances.  In Bayesian statistics the likelihood of a pixel $\boldsymbol{x}$ originating from pattern $k$ is obtained by evaluating the posterior predictive distribution (PPD) for pattern $k$. For our two-layer Gaussian mixture architecture PPDs are derived in the form of \textit{student-t} distributions by integrating out unknown mean vectors and covariance matrices of the true pattern distributions and their observed instances. This directly links observed pattern data with the hyperparameters of the model  ($\kappa_0$,$\kappa_1$,$m$,$\mu_0$, $\Sigma_0$). Optimizing hyperparameters with pixel data from the training library encodes information about observed pattern variations into the model. Technical details of the derivation of PPD for the proposed two-layer GMM are described in the supplementary material.

\subsection{Bland pixel scoring and ratioing}
To compute the likelihood of individual pixels originating from the bland pattern categories described in Section \ref{sec:lib} an ensemble version of the model discussed in Section \ref{sec:i2gmm} is used. Multiple different submodels each with different subset of channels are included in the ensemble. Ensemble models are proven to offer better generalizability and are known to be more robust with respect to noise  compared to a single model \cite{rf}. 

These likelihood scores are then used to identify denominator regions during column-wise ratioing. For a given pixel the denominator is obtained as the average spectrum of a small number of pixels with the highest bland pixel scores sharing the same column as the given pixel but lies only within $2w$ row neighborhood of that pixel, where $w$ defines the size of row neighborhood. For robust denominator-insensitive ratioing a range of $w$ values are considered to obtain multiple denominators and their corresponding ratioed spectra are averaged to obtain a single ratioed spectrum for that pixel. Once all pixels in each I/F image are ratioed this way the ratioed data are used by the pattern classifier for pixel-scale classification.

\subsection{Automated pattern classification}

Ratioed I/F data are further processed using a cascaded set of one-dimensional median filters with decreasing window sizes to gradually eliminate spikes of arbitrary heights \cite{liu2004line}.  These ratioed and despiked data are used to train the two-layer Bayesian classifier. This training process involves estimating the parameters of the PPD corresponding to each pattern. Unlike bland pixel scoring, which uses only bland pattern categories, the pattern classifier is implemented with spectral data from all patterns available in the training library. An image is classified at the pixel-scale by evaluating the likelihood of each of its pixel originating from one of the patterns in the training library and then assigning it to the pattern that maximizes this likelihood. 

\subsection{Active machine learning}

The initial training library consisted of patterns detected from a limited number of CRISM images. To obtain a more representative training library, while classifying new images, an active learning scheme is adopted. After each image is classified all detected patterns are visually inspected to confirm automated detections and training library is updated accordingly. More specifically, if a new pattern is misclassified into one of the existing patterns a new pattern class is created for this pattern in the training library. If a new spectral variant of an existing pattern is detected, the training data for that pattern is augmented with pixels from the new variant.  The classifier is retrained, i.e., PPDs are updated, every time the training data is updated. Using this active learning framework we processed over five hundred images. Our current spectral training library contains 160 patterns represented by over 400,000 spectra from 330 images. 

\section{Results}

\subsection{Diverse wallrock minerals at Jezero crater}

\begin{figure}[p]
 \centering
 \includegraphics[width=\textwidth]{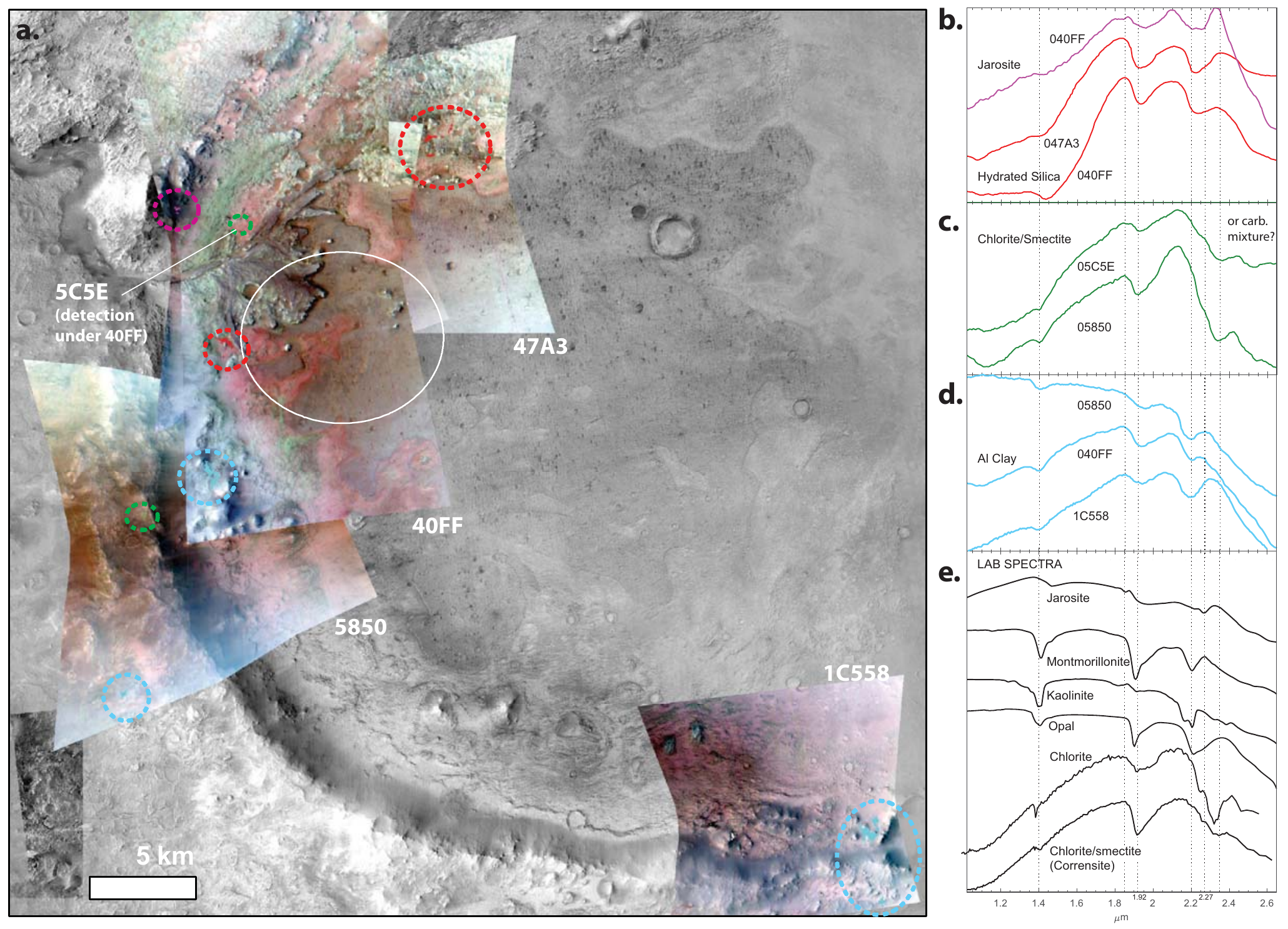}
 \caption{CRISM images covering the floors and walls of Jezero crater show sub-km exposures of Al phyllosilicates, Fe/Mg phyllosilicates (e.g. corrensite), hydrated silica, and jarosite. (a) CRISM false color images (R: 2.5 $\mu$m, G: 1.5 $\mu$m, B: 1.1 $\mu$m) overlain on a CTX mosaic. The regions of interest with colors corresponding to the spectra in (b-d) are shown, with dashed circles to flag the locations. (b-d) ratioed CRISM spectra identified by the hierarchical Bayesian algorithm. (e) library spectra from USGS \citep{usgs} and KECK/NASA reflectance experiment laboratory (RELAB).}
 \label{figureB}
  \end{figure}

Mapping of wallrock previously revealed low-Ca pyroxenes \citep{ehlmann2008orbital,ehlmann_2009a,goudge2015assessing}. Here we show also Al phyllosilicates and Fe/Mg phyllosilicates in the western wall of Jezero crater (Figure \ref{figureB}). The aluminum phyllosilicates are found on the western crater rim (FRT00005850, HRL000040FF) and the southern crater rim (FRT0001C558) at a similar elevation. The observed Al phyllosilicate spectra have an absorption centered between 2.19-2.20 $\mu$m as well as absorptions at 1.4 and 1.9 $\mu$m. The slight asymmetry in many of the spectra suggests the presence of kaolinite or another aluminum phase (Figure \ref{figureB}d). Fe/Mg phyllosilicate detections are uncommon in the walls (in contrast to other craters in the region (Ehlmann et al., 2009) but are best isolated right on the rim in FRT0005850 with 1.4, 1.9, and 2.3 $\mu$m absorptions. The long wavelength absorption is between 2.32-2.34 $\mu$m, longer than the Mg carbonates and Fe/Mg smectites that are common in Jezero sediments and basin floor deposits, and this location lacks a 2.5 $\mu$m absorption. The spectra are consistent with chlorite or mixed layer Fe/Mg smectite-chlorite phyllosilicates. Longer 2.32-2.34 $\mu$m absorptions are also found in some materials on the crater floor (e.g. in FRT0005C5E). These may be similar to the wall materials, mixed with Mg carbonates or may indicate Fe/Ca carbonates (Figure \ref{figureB}c).

\subsection{Silica and Jarosite at Jezero crater}
As also reported by \citep{tarnas2019}, we find exposures of hydrated silica within the Jezero basin (Figure \ref{figureB}). A number of small exposures $<$500$m^{2}$ are found scattered in the heavily degraded northern delta (FRT000047A3). A small exposure is also found on the southernmost lobe of the western delta (HRL000040FF, FRT00005C5E). The exposures have 1.4, 1.9, and 2.2 $\mu$m absorptions; the 2.2-$\mu$m absorption is substantially wider than in the Al-phyllosilicates (Figure \ref{figureB}b). 

In two images (HRL000040FF, FRT00005C5E) another exposure with an absorption of similar width to the hydrated silica is found, but here the band minimum is at 2.26 $\mu$m (Figure \ref{figureB}b). This suggests the presence of jarosite, separate or intermixed with the silica although at the signal to noise of the dataset, mixtures of silica with another mineral cannot be completely excluded. The location and spectral characteristics are the same in both images.

 \begin{figure}[p]
 \centering
 \includegraphics[width=\textwidth]{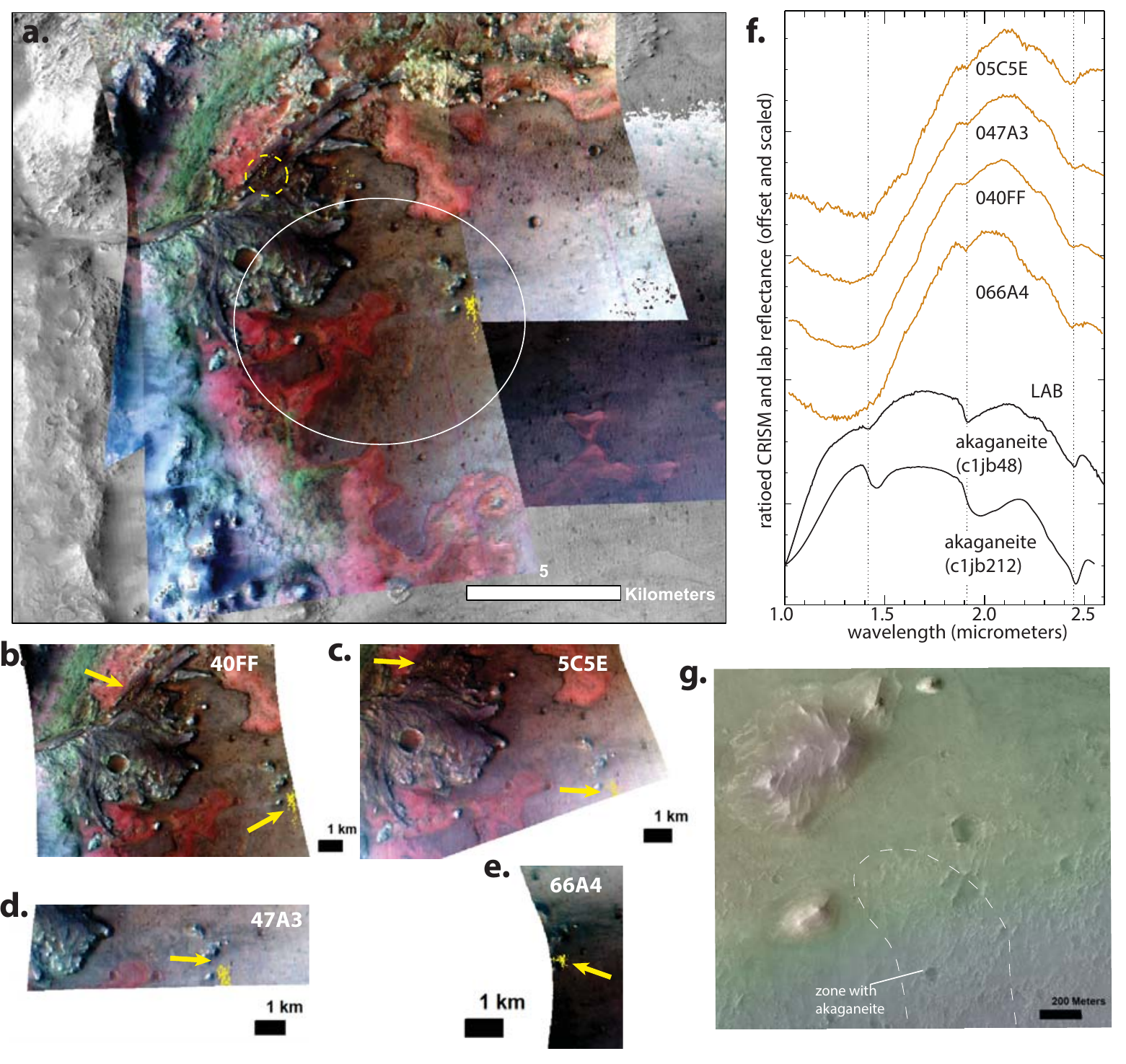}
 \caption{(a) CRISM images covering the floor of Jezero crater show akaganeite. Basemap is the same as Figure B; yellow regions indicate akaganeite, circled where the pixels are detected in multiple images. (b)-(e) zoom on segments of the CRISM images with the akaganeite sub-km exposures. (f) ratioed CRISM spectra from each of the images compared to laboratory spectra of akaganeite. (g) HiRISE digital elevation model (ESP\_023379\_1985\_\_ESP\_023524\_1985) on HiRISE showing the portion of the more rubbly floor materials with akaganeite. Elevations range from Xm to Xm.}
 \label{figureC}
  \end{figure}

\subsection{Akaganeite at Jezero crater and NE Syrtis}

A new type of hydrated mineral deposit in Jezero crater was discovered by identifying a cluster of spatially co-located but not always adjacent similar pixels by the hierarchical Bayesian model and then confirmed with traditional ratio techniques (Figure \ref{figureC}). The hydrated phase has a 1.9-$\mu$m absorption that indicates H$_2$O and a 2.45-$\mu$m absorption (Figure \ref{figureC}f). Relative to nearby spectrally "bland" materials there is also a red slope from shorter to longer wavelengths that indicates electronic transitions related to Fe mineralogy different from those of other floor materials. The spectra are most similar to akaganeite Fe$^{3+}_8$(OH,O)$_{16}$Cl$_{1.25}$·nH2O, and the spectral properties as well as geologic setting near a basin margin are similar to akaganeite reported in Sharp crater \citep{carter2015orbital}. Importantly, the phase is detected in the same locality with the same spectral characteristics in four different images (Figure \ref{figureC}b-\ref{figureC}e). The largest deposits are located near eroded remnants of deltas on the Jezero floor on the margins of a local topographic low (Figure \ref{figureC}g). The area with akaganeite appears rougher and more rubbly than surrounding floor but is otherwise geomorphologically unremarkable. 

\begin{figure}[p]
 \centering
 \includegraphics[width=\textwidth]{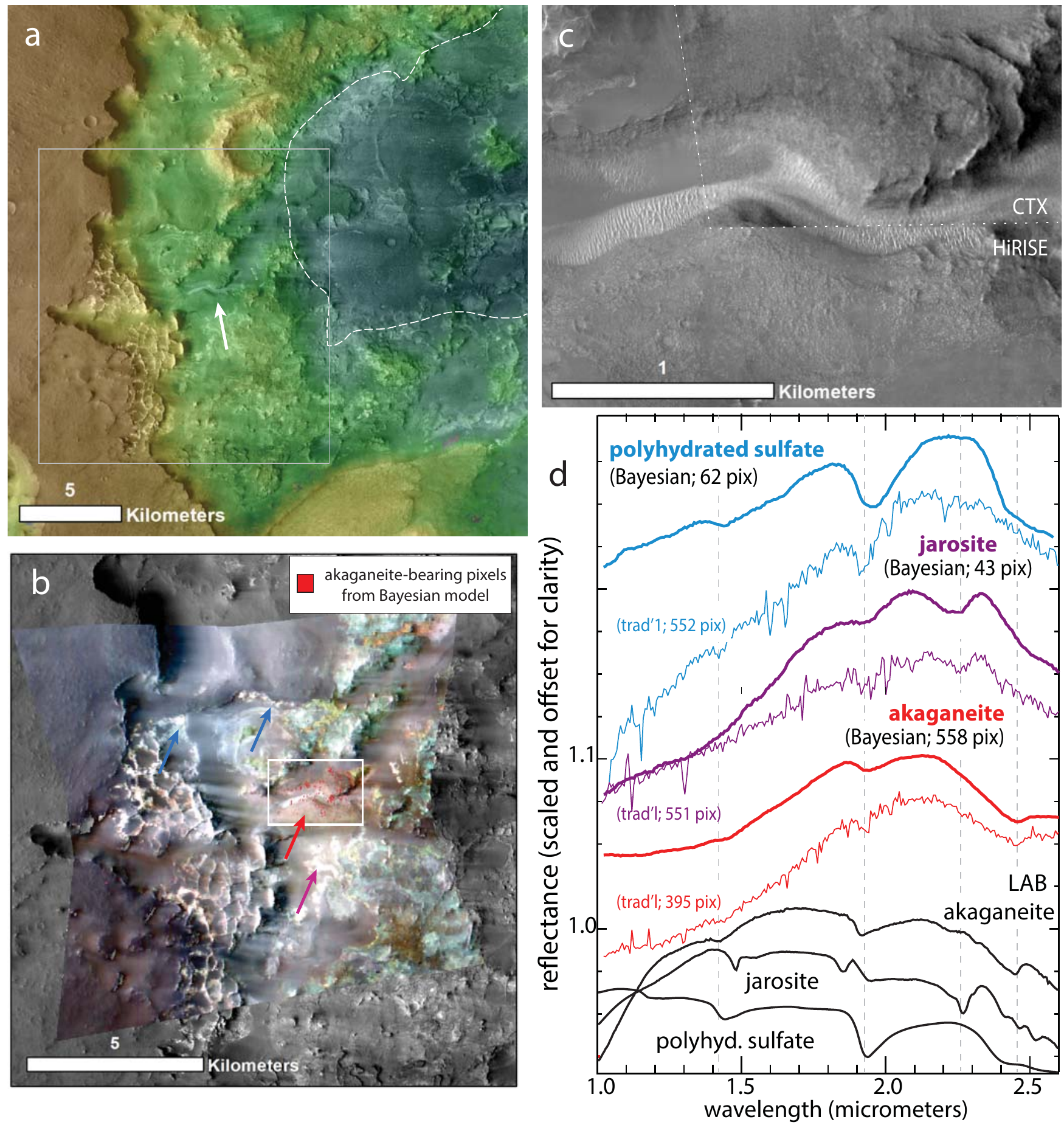}
 \caption{(a) CTX digital elevation model overlapped on a CTX mosaic from Quinn and Ehlmann (2019), showing Syrtis lavas and basin-filling deposits, incised by Late Hesperian/Early Amazonian fluvial channels (white arrow). (b) CRISM FRT00019DAA false color image (R: 2.5 $\mu$m, G: 1.5 $\mu$m, B: 1.1 $\mu$m) overlain on the CTX mosaic with pixels of akaganeite detected by a conservative threshold application of the 2-layer Gaussian Bayesian model shown in red. Arrows indicate the approximate locations of the color spectra in panel (d). (c) CTX and HiRISE images of the incised basin-filling deposits, which have the distinctive signature of akaganeite. (d) spectra of previously identified polyhydrated sulfates (blue) and jarosite (magenta) from Quinn and Ehlmann (2019) along with the new phase we identify as akaganeite (shown in comparison to library spectra in from the RELAB spectral library). The arrows in (B) signify the locations of centers of regions of interest for the spectra. The spectra from the center column obtained via the traiditional method were ratioed to the same spectral demoninator. A blue arrow to the left signifies the location of the sulfate from the Bayesian classifier. Red and magenta arrows are the sites of both traditional and Bayesian classifer-derived akaganeite and jarosite. }
 \label{figureA}
  \end{figure}

Sizeable deposits ($>$0.5 km$^2$) with an akaganeite spectral signature are also found at NE Syrtis. In CRISM image FRT00019DAA, the signature occurs in basin fill deposits that are incised by a channel that flows west to east over the Syrtis lava flows and is just upstream from late-Hesperian or early Amazonian fill deposits that host Fe/Mg phyllosilicate clay minerals (Figure \ref{figureA}; described in \citep{quinn2018pca}). The phase is spatially restricted to a specific deposit on the upstream end of the basin that has coarse layering in CRISM image FRT00019DAA (Figure \ref{figureA}c). The phase is spectrally similar to the akaganeite in Jezero but is distinct from nearby polyhydrated sulfate and jarosite spectral signatures (Figure \ref{figureA}d; e.g., \citep{ehlmann2012situ,quinn2018pca}. In addition, another deposit of akaganeite in NE Syrtis has been located using the same approach in CRISM image FRT00019538, also within basin fill deposits.

\section{Discussion}

\subsection{Two-layer Bayesian Gaussian Mixture Modeling Performance}
The proposed hierarchical Bayesian classifier improves mineral mapping in Jezero crater beyond that attained from by-hand work of previous investigators. Small exposures of uncommon phases were identified, testifying to the utility of this approach, which may lead to additional new discoveries elsewhere on Mars and offers new information for interpretation of geologic history.

\subsection{Wallrock and Jezero Floor deposits}
The wallrock of Jezero crater shares some spectral characteristics with Noachian basement materials mapped elsewhere in the regions with Fe/Mg phyllosilicates, including chlorite and smectites \citep{ehlmann_2009a,
viviano2013implications}. The Al phyllosilicate found in Jezero walls is not as typical regionally and is found at nearly the same elevation in the western and southern walls. It may be a layer of excavated basement materials, locally recording enhanced alteration, or later-formed Al phyllosilicates along the margins of the wall. The geologic context is unclear in current high resolution image data, but the signal is not   associated with the most resistant wall rock. 

Our finding of silica on Jezero crater floor units expands on similar small exposures reported previously by \citep{tarnas2019}. These may record changes in lake chemistry over time; however, their fairly limited spatial extent, which is not obviously confined to layers, may instead indicate focused zones of groundwater flow and upwelling. Sub-meter scale analysis of rock textures with Mars-2020 will differentiate between these hypotheses. 

\subsection{Environmental History Implied by Akaganeite}
This is the first report of akaganeite in the greater Nili Fossae area. Akaganeite is the best candidate to explain the observed spectral properties of this new phase discovered by the hierarchical Bayesian classifier. Longward of 1.7 –$\mu$m, the properties best, and apparently uniquely, match akaganeite. Shortward, the interpretation of Fe-related features is complicated by the fact that mafic units, which have Fe-related absorptions, serve as a denominators in ratioing to reduce artifacts.  

In both Jezero crater and NE Syrtis, the akaganeite-bearing deposits are associated with eroded, basin-filling materials formed by fluvio-lacustrine processes. This is consistent with a geologic setting where salty, Cl-bearing, Fe-bearing and possibly acidic Martian waters flowed over the southern Nili Fossae area forming a set of local lake basins, perhaps dammed by ice, which then evaporated [Skok et al., 2016; Quinn and Ehlmann, 2019]. The fluvial activity  is constrained to occur in the late Hesperian or Amazonian by superposition on the Syrtis lavas. 
The akaganeite setting in local topographic lows is similar to that of the first orbitally-detected  akaganeite in Sharp crater, also inferred to result from Fe-rich, salty waters (Carter et al., 2015).

\subsection{Implications for landed rover exploration}
At Jezero and NE Syrtis, small detections of rare phases are crucial for guiding the Mars-2020 rover and for contextualizing its discoveries. Here we are conservative in our reporting of detections, detailing only those that we were able to verify via traditional techniques, once recognized by the two-layer Bayesian approach. These encompass phases of significance for interpreting the environmental history. However, additional power for operational decision-making about the rover path could come from incorporating all detections – and their probabilities – into a systematic map of the crater, a potential subject for our future work.
\subsection{The importance of machine learning for planetary hyperspectral data analysis}
Our study demonstrates that machine learning can be highly effective in exposing tiny outcrops of rare phases in CRISM data on Mars that are not uncovered in traditional approaches to image spectroscopy data analysis. Some of these detections may offer new clues toward a more accurate and complete geologic mapping of Mars paving the way for future discoveries. Although we reported results only from select locales owing to their significance, similar outcrops of rare phases have been detected across Mars along with several interesting patterns currently being considered as candidates for new phases. Similar techniques can be applied to other imaging spectrometer data analyses for data from imaging spectrometers from other planetary bodies.

\acknowledgments
Thanks to the CRISM science and operations teams for their work to collect and process these datasets and to Jay Dickson and the Caltech Murray Laboratory for Planetary Visualization for the global CTX mosaic and other assistance with dataset registration. M.D. was sponsored by the National Science Foundation (NSF) under Grant Number IIS-1252648 (CAREER). The content is solely the responsibility of the authors and does not necessarily represent the official views of NSF. E.K.L. was supported by an NSERC PGS-D scholarship. B.L.E. thanks NASA MRO-CRISM extended mission funding for partial support. All CRISM data used in this paper are publicly available through the PDS node (http://ode.rsl.wustl.edu/mars/). Image coordinates of all detections reported in this paper are available as a supplementary file.


%
\bibliography{dundar.bib}
%




\end{document}